\documentstyle[12pt,epsf]{article}
\begin{document}
\begin{center}
{\Large \bf Exponential corrections to low-temperature 
expansion of $2D$ non-abelian models}\\

\vspace*{1cm}
{\bf O.~Borisenko\footnote{email: oleg@ap3.bitp.kiev.ua}, \
V.~Kushnir\footnote{email: kushnir@ap3.bitp.kiev.ua}}\\

\vspace*{0.5cm}
{\large \it
N.N.Bogolyubov Institute for Theoretical Physics, National Academy
of Sciences of Ukraine, 252143 Kiev, Ukraine}\\
\vspace*{0.3cm}

\end{center}
\vspace{.5cm}

\begin{abstract}
The thermodynamic limit of certain exponential corrections
to the weak coupling expansion of two-dimensional models
is investigated. The expectation values of operators contributing 
to the first order coefficient of the low-temperature expansion 
of the free energy are calculated for the order $O(e^{-\beta})$. 
They are proven to diverge logarithmically with the volume
for non-abelian models. 
\end{abstract}

It has been recently argued that the low-temperature asymptotic expansion
of two-dimensional ($2D$) non-abelian models can be non-uniform 
in the volume \cite{superinst}. As an example, authors of \cite{superinst}
involved the superinstanton boundary conditions (BC) and have shown
that the conventional perturbation theory (PT) produces results
for invariant functions which differ in the thermodynamic limit (TL) 
from those obtained with more standard BC, 
and the difference starts at $O(1/\beta^2)$.
Since the asymptotic expansion of invariant functions in the TL is unique,
the discovery of \cite{superinst} may indicate a failure of the PT. 
However, it has been shown later that the PT with superinstanton BC
diverges for higher orders and it has been argued that the low-temperature 
asymptotic expansion is non-uniform in the volume only for
superinstanton BC while there are good reasons to believe
that the PT gives true, uniform asymptotics with conventional BC
\cite{David}-\cite{QUQ} (see, however Ref.\cite{QA}).

If the low-temperature expansion 
(which is asymptotic when the volume is fixed) 
is non-uniform in the volume, one should encounter infrared (IR)
divergences (logarithmic in $2D$) either in the coefficients of 
the expansion or in the remainder to the expansion.  
It has been proven that the coefficients of the continuum PT 
done in the dimensional regularization with mass regulator term
is IR finite \cite{IR}. In this paper we assume
the same holds for bare lattice PT with periodic BC.
This assumption however cannot influence our result.
We are going to show that logarithmic divergences do appear
in the exponential corrections to the finite-volume asymptotic
expansion, already in the order $O(e^{-\beta})$. 

We work with $SU(2)$ model but to make our procedure 
clearer we perform the same calculations for the $XY$ model 
where it is known rigorously that the PT gives a uniform
asymptotic expansion \cite{XYPT}.
The arguments are very simple, both conceptually and technically.
First, let us discuss the general strategy.
In our previous papers \cite{su2}, \cite{sun} we proposed to use
an invariant link formulation to perform the low-temperature
expansion of $2D$ models. The starting point of calculations is 
the following partition function of $2D$ $SU(N)$ model 
on the periodic lattice
\begin{equation}
Z = \int \prod_l dV_l
\exp \left[ \beta \sum_l {\mbox {Re Tr}} V_l + \ln J(V) \right] \ ,
\label{lPF}
\end{equation}
\noindent
where the Jacobian $J(V)$ is a product of $SU(N)$ delta-functions
taken over all plaquettes of $2D$ lattice
\begin{equation}
J(V) = \prod_p 
\left[ \sum_r d_r \chi_r \left( \prod_{l\in p}V_l \right) \right] \ .
\label{jacob}
\end{equation}
\noindent
The sum over $r$ runs over all representations of $SU(N)$, 
$d_r=\chi_r(I)$ is the dimension of the $r$-th representation. 
The $SU(N)$ character $\chi_r$ depends on an ordered product 
of the link matrices $V_l$ along a plaquette
\begin{equation}
\prod_{l\in p}V_l = V_n(x)V_m(x+n)V_n^+(x+m)V_m^+(x) \ .
\label{prod}
\end{equation}
\noindent
The similar formulae can be written for the $XY$ model. 
In either case the formal low-temperature expansion
takes the form \cite{su2}
\begin{equation}
Z = \left [ 1+\sum_{k=1}^{\infty}\frac{1}{\beta^k}
A_k\left (\partial_h,\partial_s \right ) \right ] M(h,s) \ ,
\label{expansion}
\end{equation}
where $A_k$ are known differential operators acting on the generating
functional $M$. We find up to a constant \cite{su2}
\begin{eqnarray}
M(h,s) = \int_{-\infty}^{\infty} \prod_{x,k} d\alpha_k(x)
\int_{-\infty}^{\infty} \prod_{l,k} d\omega_k(l) 
\exp \left[ -\frac{1}{2}\omega^2_k(l) 
-i\omega_k(l)[\alpha_k(x+n) - \alpha_k(x) ] \right]  \nonumber  \\   
\times \sum_{m(x)=-\infty}^{\infty}
\exp \left[ 2\pi i\lambda\sum_xm(x)\left ( \sum_k\alpha^2_k(x)
\right )^{1/2} +
\sum_{l,k}\omega_k(l)h_k(l) + \sum_{x,k}\alpha_k(x)s_k(x) \right] \ .
\label{Gfunction}
\end{eqnarray}
\noindent
The group index $k$ takes the value $k=1,2,3$ for $SU(2)$ and
$k=1$ for the $XY$ model. The external sources $h_k(l)$ 
are coupled to the link fields $\omega_k(l)$ while $s_k(x)$ 
are coupled to the auxiliary fields $\alpha_k(x)$. In the abelian case  
$s_k(x)=0$ since the operators $A_k$ depend 
only on link fields. $\lambda = \sqrt{2\beta}$ for $SU(2)$ and 
$\lambda = \sqrt{\beta}$ for the $XY$ model.

The finite-volume asymptotic expansion arises from the configuration
$\{m_x\}=0$ for all $x$ (``vacuum'' sector). 
In \cite{sun} we have shown that the first two 
coefficients of the fixed distance correlation function 
calculated in this vacuum sector coincide with 
the conventional result. For general reasons (the uniqueness
of the asymptotic expansion in the finite volume for a given BC
and the IR finiteness of the coefficients of the expansion)
one should expect that these coefficients coincide up to an
arbitrary order. All other configurations in the sum over $m_x$
in (\ref{Gfunction}) are exponentially suppressed with $\beta$ and do not
contribute to the asymptotics. We shall however prove that
\begin{equation} 
\frac{1}{2L^2}C_{XY}^1=
\frac{1}{2L^2}A_1\left (\partial_h,\partial_s \right ) M(h,s)=
\frac{1}{32}+O(\beta e^{-\pi^2\beta}) \ ,
\label{A1XY}
\end{equation}
\noindent
for the $XY$ model (actually, this is true for any $A_k$) while
\begin{equation} 
\frac{1}{2L^2}C_{SU2}^1=
\frac{1}{2L^2}A_1\left (\partial_h,\partial_s \right ) M(h,s)=
\frac{3}{64}+O \left ( \beta \ln L e^{-2\pi^2\beta} \right ) \ ,
\label{A1SU2}
\end{equation}
\noindent
for $SU(2)$ model. The constants $1/32$ and $3/64$ are the standard 
first order coefficients of the free energy expansion.

Before we proceed to calculations we would like to present some
simple arguments why Eq.(\ref{A1SU2}) should be expected. In the vacuum 
sector the $SU(N)$ delta-function in Eq.(\ref{lPF}) reduces to the Dirac 
delta-function so that the partition function (\ref{lPF}) becomes
\begin{equation}
Z(\beta \gg 1) = \int \prod_l dV_l
\exp \left[ \beta \sum_l {\mbox {Re Tr}} V_l \right] 
\prod_{p,k}\delta (\omega_p^k) \ 
\label{lPFLT}
\end{equation}
\noindent
with the zero modes for the auxiliary fields omitted from Green functions 
\cite{su2}. $\omega_k(p)$ is a plaquette angle defined as
\begin{equation}
V_p = \prod_{l\in p}V_l =  \exp \left [ i\sigma^k\omega_k(p) \right ] 
\label{plangle}
\end{equation}
\noindent
and has the following expansion 
\begin{equation}
\omega_k(p) =  \omega^{(0)}_k(p) +
\frac{1}{\sqrt{2\beta}}\omega^{(1)}_k(p) + 
\frac{1}{2\beta}\omega^{(2)}_k(p) + \ ... \ .
\label{wkxexp}
\end{equation}
\noindent
On a dual lattice (see \cite{su2}) the delta-function 
in (\ref{lPFLT}) is expanded as
\begin{eqnarray}
\prod_{x,k}\delta (\omega_x^k) = \int_{-\infty}^{\infty}
\prod_{x,k}\frac{d\alpha_k(x)}{2\pi}
\exp \left[ -i\sum_{x,k}\alpha_k(x)\omega^{(0)}_k(x) \right] \\  \nonumber
\times \left[ 1 + \sum_{q=1}^{\infty}\frac{(-i)^q}{q!} 
\left( \sum_{x,k}\alpha_k(x) 
\sum_{n=1}^{\infty} 
\frac{\omega^{(n)}_k(x)}{(2\beta )^{n/2}} \right)^q \right] \ .
\label{DDE}
\end{eqnarray}
\noindent
If we adopt the following definitions
\begin{equation}
\omega_k(l) \to \frac{\partial}{\partial h_k(l)} \ , \ 
\alpha_k(x) \to \frac{\partial}{\partial s_k(x)} \ ,
\label{deriv}
\end{equation}
\noindent
the generating functional in the vacuum sector takes the form
\begin{equation}
M_0(h,s) = \exp 
\left[ \frac{1}{4}s_k(x)G_{x,x^{\prime}}s_k(x^{\prime}) +
\frac{i}{2}s_k(x)D_l(x)h_k(l) +
\frac{1}{4}h_k(l)G_{ll^{\prime}}h_k(l^{\prime}) \right] \ ,
\label{GFfin}
\end{equation}
\noindent
where $G_{ll^{\prime}}$ and $D_l(x^{\prime})$ are link Green functions
(see the Appendix). These functions are IR finite by construction.
It follows from the last equations that the IR finiteness of the first order
coefficient requires the finiteness of the expectation value of the
following operator
\begin{equation}
U = \frac{1}{2}\left ( \sum_{x,k}\alpha_k(x)\omega^{(1)}_k(x)
\right )^2 \ ,
\label{inffinM}
\end{equation}
\noindent
because only this operator includes IR divergent Green function 
$G_{x,x^{\prime}}$. One can show that the IR finiteness of this 
operator is equivalent to 
\begin{equation}
\langle\left (\sum_{x,k}\omega^{(1)}_k(x) \right )^2\rangle_0 = 0 \ ,
\label{inffin}
\end{equation}
\noindent
where $\langle ... \rangle_0$ refers to the following partition function
\begin{equation}
Z_0 = \int_{-\infty}^{\infty} \prod_{l,k} d\omega_k(l) 
\exp \left[ -\frac{1}{2}\omega^2_k(l) \right] 
\prod_{x,k}\delta \left (\omega^{(0)}_k(x) \right ) \ . 
\label{Z_0}
\end{equation}
\noindent
An even power of the real operator $\sum_{x,k}\omega^{(1)}_k(x)$ calculated 
over positive measure can only be zero if the operator itself
vanishes identically on all configurations contributing to (\ref{Z_0}).
Indeed, the constraint in (\ref{Z_0})
\begin{equation}
\omega^{(0)}_k(x) = 0  
\label{const0}
\end{equation}
\noindent
can be trivially solved on the original lattice in terms of site variables
\begin{eqnarray}
\omega_k(l_3)=\omega_k(x)-\omega_k(x+n) \ , \
\omega_k(l_4)=\omega_k(x+n)-\omega_k(x+n+m) \ , \nonumber  \\
\omega_k(l_1)=\omega_k(x+m)-\omega_k(x+n+m) \ , \
\omega_k(l_2)=\omega_k(x)-\omega_k(x+m) \ . 
\label{constsol}
\end{eqnarray}
\noindent
For this solution, $\sum_{x,k}\omega^{(1)}_k(x)=0$. 
In \cite{sun} we have shown how Eq.(\ref{inffin}) is fulfilled 
in terms of link Green functions (see also below).

The crucial point is that the IR finiteness of the first order 
coefficient requires the finiteness of the operator $U$
not only in the vacuum sector but for all $m_x$ in Eq.(\ref{Gfunction}).
For an arbitrary configuration $\{m_x\}$ the constraint
(\ref{const0}) becomes
\begin{equation}
\left ( \sum_k(\omega^{(0)}_k(x))^2 \right )^{1/2} = 2\pi\lambda m_x \ .  
\label{constm}
\end{equation}
\noindent
It is easy to find a non-trivial solution of (\ref{constm})
on which the operator $\sum_{x,k}\omega^{(1)}_k(x)$ does not vanish
(e.g., the ``vacuum'' solution (\ref{constsol}) for the components
$k=1,2$ and some non-trivial vortex solution for $\omega_3(l)$).
Since the integration measure is positive for any configuration
$\{m_x\}$, there is no way for Eq.(\ref{inffin}) to be satisfied
in ``non-vacuum'' sectors. This argument explains why Eq.(\ref{A1SU2}) 
is not surprising but should be expected on general grounds.

We want to prove now Eq.(\ref{A1XY}). Calculating
the generating functional (\ref{Gfunction}) for the $XY$ model
we find (sum over all repeating indices is understood)
\begin{equation}
M_{XY}(h)=e^{\frac{1}{4}h_lG_{ll^{\prime}}h_{l^{\prime}}} 
\sum_{m_x}\delta (\sum_xm_x)
e^{-\pi^2\beta m_xG_{x,x^{\prime}}m_{x^{\prime}}+
\pi\sqrt{\beta}h_lD_l(x^{\prime})m_{x^{\prime}}} \ .
\label{MXY}
\end{equation}
\noindent
Actually, Eq.(\ref{A1XY}) follows directly from this 
representation for $M_{XY}(h)$: the generating functional depends 
only on functions which are IR finite. The operator $A_1$ is given by
$\frac{1}{24}\sum_l(\partial^4/\partial h_l^4)$.
It leads to the following expansion
\begin{eqnarray}
\frac{1}{2L^2}A_1\left ( \partial_h \right ) M_{XY}(h) =  
Z_{vor}[ \frac{1}{32}-\frac{\pi^2\beta}{16L^2}\sum_{x_1x_2}D(x_1-x_2)
\langle m_{x_1}m_{x_2}\rangle_{vor} \nonumber  \\
+ \frac{\pi^4\beta^2}{48L^2}\sum_{x_1x_2x_3x_4}\sum_l
D_l(x_1)D_l(x_2)D_l(x_3)D_l(x_4)
\langle m_{x_1}m_{x_2}m_{x_3}m_{x_4}\rangle_{vor} ] \ .
\label{XYexp}
\end{eqnarray}
\noindent
Here, the expectation value $\langle ... \rangle_{vor}$ 
refers to the vortex part of the $XY$ model
\begin{equation}
Z_{vor} = \sum_{m_x}\delta (\sum_xm_x)
e^{-\pi^2\beta m_xG_{x,x^{\prime}}m_{x^{\prime}}} \ .
\label{Zvor}
\end{equation}
\noindent
The first exponential correction comes from the configuration
$m_x=\delta_{xy}-\delta_{xz}$. We find 
\begin{equation}
\frac{1}{L^2}\sum_{x_1x_2}D(x_1-x_2)\langle m_{x_1}m_{x_2}\rangle_{vor}=
\frac{2}{L^2}\sum_{y\ne z}D(y-z)e^{-2\pi^2\beta D(y-z)}
\left [ 1+O(e^{-\beta}) \right ] \ ,
\label{exampl}
\end{equation}
\noindent
and similar expression can be written for the last term in (\ref{XYexp}).
Since $D(y-z)\geq 1/2$ for $y\ne z$, Eq.(\ref{A1XY}) follows.

Let us now turn to the $SU(2)$ model. 
It seems to be a difficult task to calculate
the generating functional for an arbitrary configuration 
$\{ m_x \}$ because of the square root in (\ref{Gfunction}). 
We therefore restrict ourself to the first 
exponential correction which again arises from 
$m_x=\delta_{xy}-\delta_{xz}$, i.e. we write
$$
\sum_{m(x)=-\infty}^{\infty}\exp [ 2\pi i\lambda\sum_xm(x)\alpha (x) ]
=1+4\sum_{y\ne z}\cos (2\pi\lambda\alpha (y))\cos (2\pi\lambda\alpha (z))
+ ...  \ .
$$
Using the inverse Laplace transform
$$
\cos (2a^{1/2}y^{1/2})=\frac{1}{2i}\left (\frac{y}{\pi} \right )^{1/2}
\int_{C-i\infty}^{C+i\infty}\frac{dt}{\sqrt{t}} e^{yt-a/t} \ , \
C > 0 \ ,
$$
the generating functional can be reduced to the form
\begin{equation}
M_{SU(2)}(h,s)=M_0(h,s)\left ( 1 + 4\sum_{y\ne z} M_1(h,s) 
+ O(e^{-4\pi^2\beta}) \right ) \ ,
\label{MSU2}
\end{equation}
\noindent
where $M_0$ is given in (\ref{GFfin}). An integral
form of $M_1$ reads ($G_0 \equiv G_{x,x}$)
\begin{eqnarray}
M_1(h,s)=-\frac{\pi\beta}{2G_{yz}}e^{-4\pi^2\beta G_0}
\int_{C-i\infty}^{C+i\infty}\frac{dtds}{(ts-1)^{3/2}}
(tG_{yz}-G_0)(sG_{yz}-G_0) \nonumber  \\
\exp \left [ 2\pi^2\beta G_{yz} (t+s)+\frac{1}{4G_{yz}(ts-1)}
(ta_k^2(z)+sa_k^2(y)-2a_k(z)a_k(y)) \right ] \ .
\label{M1SU2}
\end{eqnarray}
\noindent
We use the notation ($h_k(l) \equiv h_k(n;x)$)
$$
a_k(y)=\sum_{x}G_{xy}\left [ h_k(x)-is_k(x) \right ] \ , \
h_k(x)=\sum_{n=1}^2 \left [ h_k(n;x)-h_k(n;x-n) \right ] \ .
$$
As an example, let us give an expression for $M_1$ at vanishing
sources
\begin{equation}
M_1(0,0)=\frac{1}{2}e^{-4\pi^2\beta D(y-z)}
\left (1- \frac{4\pi^2\beta}{G_{yz}}D^2(y-z) \right ) +
O\left ( e^{-8\pi^2\beta G_0} \right ) \ .
\label{M1PFTL}
\end{equation}
\noindent
One sees that the convergence to the TL is very slow, like
$O(1/\ln L)$. 

The proof of Eq.(\ref{A1SU2}) is now straightforward. 
The operator $A_1$ reads \cite{su2}
\begin{equation}
A_1=\frac{1}{24}\sum_l\left ( \sum_k\omega_k^2(l) \right )^2-
\frac{1}{3}\sum_l\left ( \sum_k\omega_k^2(l) \right ) -
i\sum_x\sum_k\alpha_k(x)\omega_k^{(2)}(x) - U \ ,
\label{A1def}
\end{equation}
\noindent
where $U$ is given in (\ref{inffinM}).
In \cite{su2} we have shown that
\begin{equation}
\frac{1}{2L^2}A_1\left (\partial_h,\partial_s \right ) M_0(h,s)
=\frac{3}{64} \ .
\label{A1zero}
\end{equation}
\noindent
It is seen from (\ref{A1def}) that the IR divergent 
function $G_{x,x^{\prime}}$ can arise only from the third and fourth terms.
It is easy to prove that the operator
$\sum_x\sum_k\alpha_k(x)\omega_k^{(2)}(x)$ is IR finite.
Therefore, it is sufficient to study the contribution
from the last term in (\ref{A1def}). We find
\begin{equation}
\frac{1}{2L^2}U\left (\partial_h,\partial_s \right ) 
M_{SU(2)}(h,s)=U_{div}+U_{fin} \ .
\label{Uexp}
\end{equation} 
\noindent
$U_{fin}$ denotes a finite contribution which depends only on IR finite link
Green functions. For $U_{div}$ we find (up to the terms vanishing in the TL)
\begin{equation}
U_{div}=\frac{1}{2L^2}\sum_{x,x^{\prime}}G_{x,x^{\prime}}
\left [ \frac{3}{16}W_{x,x^{\prime}}^{(0)} +
5\pi^2\beta\sum_{yz}e^{-4\pi^2\beta D(y-z)} 
W_{x,x^{\prime}}^{(1)}(y,z) \right ] \ ,
\label{Udiv}
\end{equation} 
\noindent
where
\begin{equation}
W_{x,x^{\prime}}^{(0)}=\sum_{i<j}^4\sum_{i^{\prime}<j^{\prime}}^4
(G_{l_i,l^{\prime}_{i^{\prime}}}G_{l_j,l^{\prime}_{j^{\prime}}} -
G_{l_i,l^{\prime}_{j^{\prime}}}G_{l_j,l^{\prime}_{i^{\prime}}}) \ ,
\label{W0}
\end{equation} 
\noindent
\begin{equation}
W_{x,x^{\prime}}^{(1)}(y,z)=\sum_{i<j}^4\sum_{i^{\prime}<j^{\prime}}^4
(G_{l_i,l^{\prime}_{i^{\prime}}}B_{l_j,l^{\prime}_{j^{\prime}}} -
G_{l_i,l^{\prime}_{j^{\prime}}}B_{l_j,l^{\prime}_{i^{\prime}}}+
G_{l_j,l^{\prime}_{j^{\prime}}}B_{l_i,l^{\prime}_{i^{\prime}}} -
G_{l_j,l^{\prime}_{i^{\prime}}}B_{l_i,l^{\prime}_{j^{\prime}}}) \ ,
\label{W1}
\end{equation} 
\noindent
\begin{equation}
B_{l,l^{\prime}}(y,z)=\left ( D_l(y)-D_l(z) \right )
\left ( D_{l^{\prime}}(y)-D_{l^{\prime}}(z) \right ) \ .
\label{Bdef}
\end{equation} 
\noindent
The link $l_i$ ($l^{\prime}_{j^{\prime}}$) refers to one of four
links attached to a given site $x$ ($x^{\prime}$).
Let us substitute $G_{x,x^{\prime}}=G_0-D(x-x^{\prime})$ in (\ref{Udiv}). 
Since the function $D(x-x^{\prime})$ is IR finite we conclude that the IR 
finiteness of Eq.(\ref{Udiv}) requires
\begin{equation}
\sum_{x,x^{\prime}}W_{x,x^{\prime}}^{(0)}=0 \ , \ 
\sum_{x,x^{\prime}}W_{x,x^{\prime}}^{(1)}(y,z)=0 \ .
\label{IRreq0}
\end{equation}
\noindent
The first of these equations is nothing but another form of 
Eq.(\ref{inffin}). Its validity can be proven directly.
Using (\ref{orthog1}) we rewrite (\ref{IRreq0}) as
\begin{equation}
\sum_{x,x^{\prime}}W_{x,x^{\prime}}^{(0)}=\frac{1}{4}\sum_{b_1b_2}
A(b_1b_2)\left [ A(b_1b_2)- A(b_2b_1) \right ] \ ,
\label{W0rep}
\end{equation}
\noindent
where sums over $b_i$ run over all links of the lattice and
$$
A(b_1b_2) = \sum_x\sum_{i<j}G_{b_1l_i}G_{b_2l_j} \ .
$$
The equality $A(b_1b_2)- A(b_2b_1)=0$ follows from 
(\ref{Glleq}) and (\ref{Gllsum}). This equality ensures
the IR finiteness of the first order coefficient in the vacuum sector.

However, the second equation in (\ref{IRreq0}) cannot be satisfied,
in accordance with our general arguments. Indeed, using 
the same tricks as above we get
\begin{equation}
\sum_{x,x^{\prime}}W_{x,x^{\prime}}^{(1)}(y,z)=\frac{1}{2}
\sum_b \left (F_b(y)-F_b(z)  \right )^2 \ ,
\label{W1rep}
\end{equation}
\noindent
where
\begin{equation}
F_b(y) = \sum_x\sum_{i<j}\left ( G_{bl_i}D_{l_j}(y)-
G_{bl_j}D_{l_i}(y) \right ) \ .
\label{Fby}
\end{equation}
\noindent
It follows from the last two equations that in order to ensure the IR 
finiteness, the function $F_b(y)$ must be $y$-independent. In fact, 
$F_b(y)$ does depend on $y$ as it follows from the definition
of functions $G_{bl}$ and $D_{l}(y)$. To see this dependence directly,
one can use Eqs.(\ref{Glleq}), (\ref{Dleq}) (see the Appendix)
for the link Green functions and perform calculations in momentum 
space. From (\ref{W1rep}) it also follows that 
$\sum_{x,x^{\prime}}W_{x,x^{\prime}}^{(1)}(y,z)$ depends only
on a difference $(y-z)$. Since $W_{x,x^{\prime}}^{(1)}(y,y)=0$
we end up with (omitting finite terms)
\begin{equation}
\frac{1}{2L^2}\sum_{x,x^{\prime}}G_{x,x^{\prime}}
\sum_{yz}e^{-4\pi^2\beta D(y-z)} W_{x,x^{\prime}}^{(1)}(y,z) =
G_0e^{-2\pi^2\beta}\sum_{x,x^{\prime}}
W_{x,x^{\prime}}^{(1)}(0,1) + O(e^{-c\pi^2\beta}) , \ 
\label{res}
\end{equation} 
\noindent
where $c>2$. Since $G_0\approx \frac{1}{\pi}\ln L$, this completes 
the proof of Eq.(\ref{A1SU2}).

We have also computed the second term in (\ref{Udiv})
numerically. For $\beta =2$ we find the values
$3.036\cdot 10^{-15}$ for $L=200$, $3.267\cdot 10^{-15}$ for $L=300$ 
and $3.558\cdot 10^{-15}$ for $L=500$.
These numbers are so small compared to the first order coefficient
$3/128$ (at $\beta =2$) that even if there are 
deviations from the conventional PT in the TL, it seems there is no
chance to see them in the Monte-Carlo simulations.

Let us add some comments. 

I. $2D$ $SU(N)$ principal chiral models are essentially compact models
whose Gibbs measure obeys a certain periodicity required by the
invariant group measure. As a rule, the weak coupling
expansion needs a construction of the Gaussian ensemble
which destroys the compactness and periodicity of the original
measure. When the volume is fixed 
the corrections coming from the extension of the integration
region to infinity and from the breaking of periodicity
are exponentially small and thus do not contribute to the
asymptotic expansion. The subtle question is what happens
with these corrections in the large volume limit, in particular
whether they are infrared finite or not. In our opinion, the link formulation 
of these models presents a good tool for resolving this problem.
Summation over $m_x$ in the expression for the generating functional
(\ref{Gfunction}) appears due to using group invariant delta-function.
In the abelian case it corresponds to the summation over vortex
configurations. Probably, some similar interpretation can be given in
the $SU(2)$ case (it is, of course, also tempting to speculate that
these configurations are responsible for a mass gap generation
similar to vortices in the $XY$ model).   
In this paper we have investigated certain exponential corrections
to the first order coefficient in the expansion of the free energy
which appear from the mentioned summation. Our main result, 
Eq.(\ref{A1SU2}), shows that the IR divergences do not cancel for 
non-abelian models. 

However since such divergences appear only in exponentially suppressed
terms one could ask how significant they are, in particular 
for the construction of the continuum limit.
One considers asymptotics obtained by letting $L$ go to infinity
as a certain function of $\beta$, e.g. for the XY model one takes
$L=O(\exp\sqrt{\beta})$. The proof of the asymptoticity then
follows from certain inequalities and the fact that the PT
coefficients with Dirichlet and free BC differ only
by exponentially small corrections for sufficiently large $\beta$
(see section 4 of \cite{XYPT}). A strategy of a similar proof for
non-abelian models was outlined in \cite{QUQ}. 
Since one can take a power dependence on $\beta$ for $L$,
the second term on the right-hand side of Eq.(\ref{A1SU2}) remains
properly bounded. If however one is interested in the limit $L\to\infty$
where $L$ is $\beta$-independent, we conclude that the coefficients
of the low-temperature expansion are not bounded uniformly in $L$. 

II. The second type of exponential corrections comes from 
the restriction of the integration region over link variables.
For example, the one-link integral in the $XY$ model is given by
\begin{eqnarray}
I_l=\int_{-\pi\sqrt{\beta}}^{\pi\sqrt{\beta}}d\omega (l) 
\exp \left[ -\frac{1}{2}\omega^2(l) 
-i\omega (l)[\alpha (x+n) - \alpha (x) ] \right]  \nonumber  \\
= (2\pi)^{1/2}e^{-\frac{1}{2}(\alpha (x+n) - \alpha (x))^2}
\left (\mbox{erf}\left ( \pi\sqrt{\beta /2}+\frac{i}{\sqrt{2}}
(\alpha (x+n) - \alpha (x))\right ) + c.c. \right) \ ,
\label{onelink}
\end{eqnarray}
\noindent
where $\mbox{erf}(z)$ is the error function. From here 
one derives uniform bounds both for $M_{XY}(h=0)$, for the 
coefficients $A_k$ and for the correlation function \cite{bounds}.
Our result actually shows that such uniform 
bounds in $L$ do not exist for the coefficients of the asymptotic 
expansion of the non-abelian model, if $L$ is $\beta$-independent.

\vglue 0.5cm
\begin{center}
{\Large \bf Appendix }
\end{center}
\vglue 0.3cm

The functions $G_{ll^{\prime}}$ and $D_l(x^{\prime})$
are defined as
\begin{equation}
G_{ll^{\prime}} = 2\delta_{l,l^{\prime}} - G_{x,x^{\prime}} -
G_{x+n,x^{\prime}+n^{\prime}} + G_{x,x^{\prime}+n^{\prime}} + 
G_{x+n,x^{\prime}} \ ,
\label{Gll12D}
\end{equation}
\noindent
\begin{equation}
D_l(x^{\prime}) = G_{x,x^{\prime}} - G_{x+n,x^{\prime}} =
D(x-x^{\prime}+n)-D(x-x^{\prime}) \ , \ l=(x,n) \ , 
\label{Dxl2D}
\end{equation}
\noindent
where
\begin{equation}
D(x-y) = G_0-G_{x,y} = \frac{1}{L^2} \sum_{k_n=0}^{L-1}
\frac{1 - e^{\frac{2\pi i}{L}k_n (x_n-y_n)}}
{2-\sum_{n=1}^2\cos \frac{2\pi}{L}k_n} \ , \ k_n^2\ne 0 .
\label{Dx}
\end{equation}
\noindent
They satisfy the following equations
\begin{equation}
G_{l_1l^{\prime}}+G_{l_2l^{\prime}}-G_{l_3l^{\prime}}-
G_{l_4l^{\prime}} = 0 \ ,
\label{Glleq}
\end{equation}
\noindent
\begin{equation}
D_{l_1}(x^{\prime})+D_{l_2}(x^{\prime})-
D_{l_3}(x^{\prime})-D_{l_4}(x^{\prime})=2\delta_{x,x^{\prime}} \ ,
\label{Dleq}
\end{equation}
\noindent
\begin{equation}
\sum_x\left ( G_{l_1b_1}G_{l_2b_2}+G_{l_3b_1}G_{l_4b_2} - 
G_{l_1b_2}G_{l_2b_1}-G_{l_3b_2}G_{l_4b_1} \right ) = 0 \ ,
\label{Gllsum}
\end{equation}
\noindent
where four links $l_i$ are attached to a given site $x$.
One proves the following ``orthogonality'' relations for 
the link functions 
\begin{eqnarray}
\label{orthog1}
\sum_{b}G_{lb}G_{bl^{\prime}}=2G_{ll^{\prime}} \ ,  \
\sum_{b}D_b(x)G_{bl^{\prime}}=0 \ , \\
\label{orthog}
\sum_{b}D_b(x)D_b(x^{\prime})=2G_{x,x^{\prime}} \ ,
\end{eqnarray}
\noindent
where the sum over $b$ runs over all links of $2D$ lattice.


\begin{thebibliography}{99}

%
\bibitem{superinst} A.~Patrascioiu, E.~Seiler, Phys.Rev.Lett.
74 (1995) 1920.
%
\bibitem{David} F.~David, Phys.Rev.Lett. 75 (1995) 2626.
%
\bibitem{QUQ} F.~Niedermayer, M.~Niedermaier and P.~Weisz, 
hep-lat/9612002.
%
\bibitem{QA} A.~Patrascioiu, E.~Seiler, Phys.Rev. D57 (1998) 1394.
%
\bibitem{IR} F.~David, Phys.Lett. B96 (1980) 371.
%
\bibitem{XYPT} J.~Bricmont, J.-R.~Fontaine, J.L.~Lebowitz, E.H.~Lieb, 
and \ T.~Spencer, Comm.Math.Phys. 78 (1981) 545.
%
\bibitem{su2} O.~Borisenko, V.~Kushnir, A.~Velytsky, hep-lat/9809133.
%
\bibitem{sun} O.~Borisenko, V.~Kushnir, hep-lat/9905025.
%
\bibitem{bounds} O.~Borisenko, V.~Kushnir, in preparation.
%

\end{thebibliography}
\end{document}